\documentclass[aps,prl,twocolumn,showpacs,preprintnumbers,amsmath,amssymb,refmerge,superscriptaddress]{revtex4}
\usepackage{graphicx}
\usepackage{dcolumn}
\usepackage{longtable}
\def\pr{{Phys. Rev.}~}
\def\prl{{ Phys. Rev. Lett.}~}
\def\pl{{ Phys. Lett.}~}

\newcommand{\ra}{\rightarrow}

\begin{document}

\title{Observation of the decay $\overline{B}{}^0 \longrightarrow D_s^+ \Lambda \overline{p}$}

  \author{T.~Medvedeva}\affiliation{Institute for Theoretical and Experimental Physics, Moscow} 
  \author{R.~Chistov}\affiliation{Institute for Theoretical and Experimental Physics, Moscow} 
  \author{K.~Abe}\affiliation{High Energy Accelerator Research Organization (KEK), Tsukuba} 
  \author{K.~Abe}\affiliation{Tohoku Gakuin University, Tagajo} 
  \author{I.~Adachi}\affiliation{High Energy Accelerator Research Organization (KEK), Tsukuba} 
  \author{H.~Aihara}\affiliation{Department of Physics, University of Tokyo, Tokyo} 
  \author{D.~Anipko}\affiliation{Budker Institute of Nuclear Physics, Novosibirsk} 
  \author{T.~Aushev}\affiliation{Swiss Federal Institute of Technology of Lausanne, EPFL, Lausanne}\affiliation{Institute for Theoretical and Experimental Physics, Moscow} 
  \author{A.~M.~Bakich}\affiliation{University of Sydney, Sydney NSW} 
  \author{V.~Balagura}\affiliation{Institute for Theoretical and Experimental Physics, Moscow} 
% ...

\collaboration{The Belle Collaboration}
 
\begin{abstract}

We report the first observation of the decay $\overline{B}{}^0 \rightarrow D_s^+ \Lambda \overline{p}$ with a statistical significance of $6.6\,\sigma$. 
We measure ${\cal B}(\overline{B}{}^0 \rightarrow D_s^+ \Lambda \overline{p}) = (2.9 \pm 0.7\pm 0.5 \pm 0.4)\times 10^{-5}$, where the first error is statistical, the second is systematic and the third error comes from the uncertainty in ${\cal B}(D_s^+\rightarrow\phi\pi^+)$. 
The data used for this analysis was accumulated at the 
$\Upsilon(4S)$ resonance, using the Belle detector at the 
KEKB asymmetric-energy $e^+e^-$ collider.
The integrated luminosity of the data sample is 
$414\,~\mathrm{fb}^{-1}$,
corresponding to $449\times 10^{6}$ $B{\bar B}$ pairs.

\end{abstract}


\maketitle

{\renewcommand{\thefootnote}{\fnsymbol{footnote}}}
\setcounter{footnote}{0}
In the past few years, new measurements of baryonic $B$ meson decays by 
Belle \cite{BELLE1, BELLE2, BELLE3, BELLE4, BELLE5}
and CLEO \cite{CLEO1, CLEO2} have revived experimental \cite{BELLE6, BELLE7, BELLE8} 
and theoretical interest \cite{HOU1, HOU3, ROSNER, KERBIKOV, HAIDEN, ENTEM} in
such processes. Multi-body baryonic decay 
modes are found to have larger branching fractions than two-body modes, 
and the baryon-pair invariant mass spectrum peaks near threshold in the case
of multi-body decays \cite{THRESHOLD}. This feature was conjectured in Ref. \cite{HOU2}. Further 
investigations of the Dalitz plot \cite{DALITZ} and the angular correlations for events in the threshold region \cite{BELLE9} offer better understanding of the underlying dynamics.

To date, nothing is known experimentally about charmful baryonic $B$ decays 
with the creation of an $s\bar{s}$ pair. 
$\overline{B}{}^0$ mesons can decay to $D_s^+ \Lambda \overline{p}$ through 
the Cabibbo favoured $b\rightarrow c\overline{u}d$ process. They can also 
decay to the charge conjugate final state through the Cabibbo suppressed
$b\rightarrow u\overline{c}d$ process, opening a new avenue for future
$CP$ asymmetry studies.
We report here the first observation of the decay $\overline{B}{}^0\rightarrow D_s^+ \Lambda \overline{p}$ using
$414\,{\rm fb}^{-1}$ of data, corresponding to $449 \times 10^6$ $B\overline{B}$ 
pairs, collected at the $\Upsilon(4S)$ resonance with the Belle detector at the KEKB asymmetric-energy $e^+e^-$
collider~\cite{kekb}.
Since the $D_s^+ \Lambda\overline{p}$ final state may get a contribution from the
$D^0 p \rightarrow D_s^+ \Lambda$ final state rescattering, the previously observed
$B^0 \rightarrow D^0 p\overline{p}$ decay \cite{BELLE5} could be one of the sources for the $D_s^+ \Lambda\overline{p}$ final state.
Inclusion of charge conjugate states is implicit throughout this paper.

The Belle detector is a large-solid-angle magnetic spectrometer that consists of a silicon
vertex detector (SVD), a 50-layer central drift chamber (CDC), an array of aerogel threshold
Cherenkov counters (ACC), a barrel-like arrangement of time-of-flight scintillation counters
(TOF), and an electromagnetic calorimeter (ECL) comprised of CsI(Tl) crystals located
inside a superconducting solenoid coil that provides a 1.5 T magnetic field. An iron flux-return located outside the coil is instrumented to detect $K^0_L$ mesons and to identify muons
(KLM). The detector is described in detail elsewhere~\cite{belle_detector}. 
Two different inner detector configurations were used. 
For the first sample of 152 million $B\bar{B}$ pairs, a 2.0\,cm
radius beampipe and a 3-layer silicon vertex detector were used; for the latter 297 million
$B\bar{B}$ pairs, a 1.5\,cm radius beampipe, a 4-layer silicon detector and a small-cell inner
drift chamber were used~\cite{svd2}.
We use a GEANT-based Monte Carlo (MC) simulation to model 
the response of the detector and determine the efficiency~\cite{montecarlo}.

Pions, kaons and protons are identified 
using a likelihood ratio method, which combines information 
from the TOF system and ACC counters with $dE/dx$ measurements using the CDC~\cite{ID}.

In this analysis we reconstruct $D^+_s$ candidates by using $D_s^+ \rightarrow \phi \pi^+$, $\overline{K}{}^{*0} K^+$ and $K^0_S K^+$ decay modes. Candidate $\Lambda$ baryons are reconstructed via the $\Lambda \rightarrow p \pi^-$ decay.

For $\Lambda$ hyperons we require an invariant mass within $\pm3$ MeV$/c^2$ of the nominal $\Lambda$ mass \cite{PDG}. 
The distance between the $\Lambda$ decay vertex position and beam interaction point (IP) in the $r - \phi$ plane, 
$dr(\Lambda)$, is required to be greater than 0.5\,cm. The angle $\alpha_{\Lambda}$, between the $\Lambda$ momentum vector and the vector pointing from the IP to the decay vertex, must satisfy $\cos \alpha_{\Lambda}>0.95$. We also require $dz(\Lambda)<0.5$\,cm, 
where $dz(\Lambda)$ is the difference in the $z$-coordinates (the $z$ axis is parallel to the $e^+$ beam) between the $\pi$ and $p$ tracks at vertex position.
We reconstruct neutral kaons via the decay $K_S^0\rightarrow\pi^+\pi^-$ and
require its invariant mass to be within $\pm$10 MeV$/c^2$ (about 4\,$\sigma$) of the nominal $K_S^0$ mass. We also require $dz(K_S^0) < 1$\,cm, $dr(K_S^0)>0.01$\,cm and $|\cos (\alpha_{K_S^0})|>0.95$, where $dz$, $dr$ and $\alpha$ are defined in a way similar to the case of the $\Lambda$ hyperon.

 We use a mass and vertex constrained fit for $D_s^+ \rightarrow K^+ K^- \pi^+$,
and require the $\phi$ invariant mass to be within $\pm$10\,MeV$/c^2$ 
and $\overline{K}{}^{*0}$ invariant mass within $\pm$50\,MeV/$c^2$ of the nominal 
masses for the $D_s^+ \rightarrow \phi \pi^+$ and $D_s^+ \rightarrow \overline{K}{}^{*0} K^+$, respectively.
 Finally, we apply helicity requirements: $|\cos {\Theta_{\phi}}|>0.3$ and  $|\cos {\Theta_{\overline{K}{}^{*0}}}|>0.3$ for $D_s^+ \rightarrow \phi \pi$ and $D_s^+ \rightarrow \overline{K}{}^{*0}K^+$, respectively. The helicity angle $\Theta_{\phi(K^*)}$ is defined as the angle between the $K (\pi)$ meson momentum and the $D_s$ meson momentum in the $\phi $ $(K^*)$ rest frame. For $D_s$ candidates we use a mass window that extends  $\pm 15$\,MeV/$c^2$ around the nominal $D_s$ mass value.
We use a large sample of inclusive $\Lambda$ and $D_s$ signals, applying the selections described above, to verify that their mass peaks are well described by two Gaussians, corresponding to the core and the tail of the distribution, where the tail fraction is 35 to 50\,\%. The signal mass windows that are used in the analysis correspond to approximately $4\,\sigma$ for the core and $2\,\sigma$ for the tail Gaussian. For the inclusive signals data and MC agree.

 To suppress the continuum background ($e^+e^- \rightarrow q \overline{q}$, where $q = u, d, s, c$), we require the ratio of the second to zeroth Fox-Wolfram moment \cite{fox} to be less than 0.5. We also require the cosine of the reconstructed $B$ meson direction with respect to the $z$-axis in center-of-mass (c.\,m.) frame, $|\cos \theta_B|$, to be less than 0.8.

The $B$ candidates are identified by their mass difference, $\Delta M = M(B) - m_B$ and their beam-energy constrained mass, $M_{\rm bc}=\sqrt{E_{\rm beam}^2 - (\sum_i \vec{p}_i)^2}$, where $E_{\rm beam} = \sqrt{s}/2$ is the beam energy
 and $\vec{p}_i$ are three-momenta of the $B$ candidate decay products in the c.\,m. system, $M(B)$ is the reconstructed mass of the $B$ candidate and $m_B$ is the world average $B$ meson mass.
 We do not use the widely applied kinematic variable $\Delta E = E_B - E_{\rm beam}$, where $E_B$ is the energy of the reconstructed $B$ in the c.\,m. system, since $\Delta E$ has a large correlation with $M_{\rm bc}$ for signal due to the small energy release in the decay under study. By contrast, $\Delta M$ and $M_{\rm bc}$ are uncorrelated\,\cite{DM1, DM2} as confirmed by MC.
We select $B$ candidates with $M_{\rm bc}>5.2$\,GeV/$c^2$ and $|\Delta M|<0.2$\,GeV/$c^2$.

\begin{table*}[t]
\caption{Summary of the fit results, efficiencies, statistical significances and branching fractions obtained from the 2D $\Delta M-M_{\rm bc}$ fit.}
\begin{center}
\begin{tabular}{c|c|c|c|c}
\hline
~~~~~Decay Mode & ~~~~~~Yield~~~~~~ & ~~~~~~Efficiency ($10^{-4}$)~~~~~~ & ~~~~~~Significance~~~~~~ & ~~~~~~$\cal B$ ($10^{-5}$)~~~~~~\\
\hline
\hline
$\overline{B}{}^0 \ra D_s^+ \Lambda \overline{p}$, $D_s^+\ra \phi \pi^+$ & $6.5 \pm 2.6$ &4.90 & 4.7$\sigma$ & $3.0 \pm 1.2$ \\
$\overline{B}{}^0 \ra D_s^+ \Lambda \overline{p}$, $D_s^+\ra\overline{K}{}^{*0} K^+$ & $4.0 \pm 2.5$ &4.31 & 2.3$\sigma$ & $2.1 \pm 1.3$ \\
$\overline{B}{}^0 \ra D_s^+ \Lambda \overline{p}$, $D_s^+\ra K_S^0 K^+$  & $7.9 \pm 3.1$ &4.83 & 4.2$\sigma$ & $3.6 \pm 1.4$ \\
\hline
~~~$\overline{B}{}^0 \ra D_s^+ \Lambda \overline{p}$, simultaneous fit~~~ & & & 6.6$\sigma$ & $2.9 \pm 0.7$ \\
\hline
\hline
\end{tabular}\end{center}
\label{2D}
\end{table*}

\begin{figure}[htb]
\centering
\includegraphics[width=0.5\textwidth]{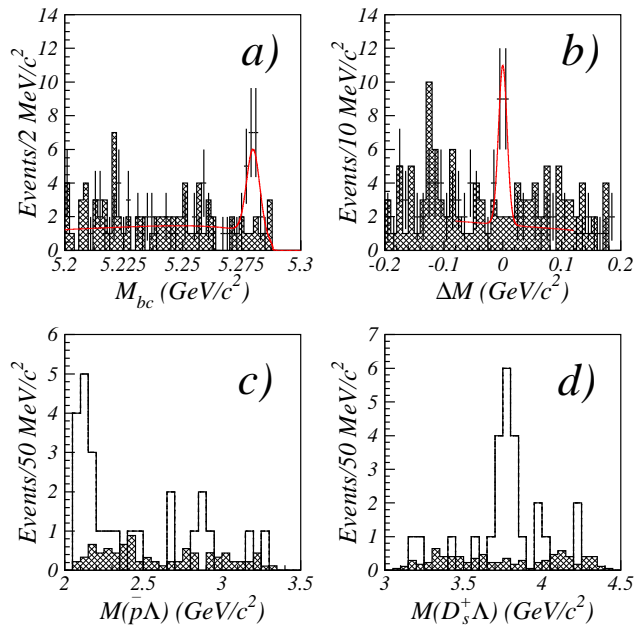}
\caption{ 
(a) The $M_{\rm bc}$ and (b) $\Delta M$ distributions for the  $\overline{B}{}^0 \ra D_s^+ \Lambda \overline{p}$ candidates (triangles with error bars). The hatched histograms show the $D_s^+$ mass sidebands normalized to the signal region. 
The overlaid curves are fit results (see text for details). 
(c) The $\overline{p}\Lambda$ and (d) $D^+_s\Lambda$ invariant mass distributions in the $B$-signal region (open histogram) and in the $B$-sideband (hatched histogram).
}
\label{2}
\end{figure}

\begin{figure}[htb]
\centering
\includegraphics[width=0.5\textwidth]{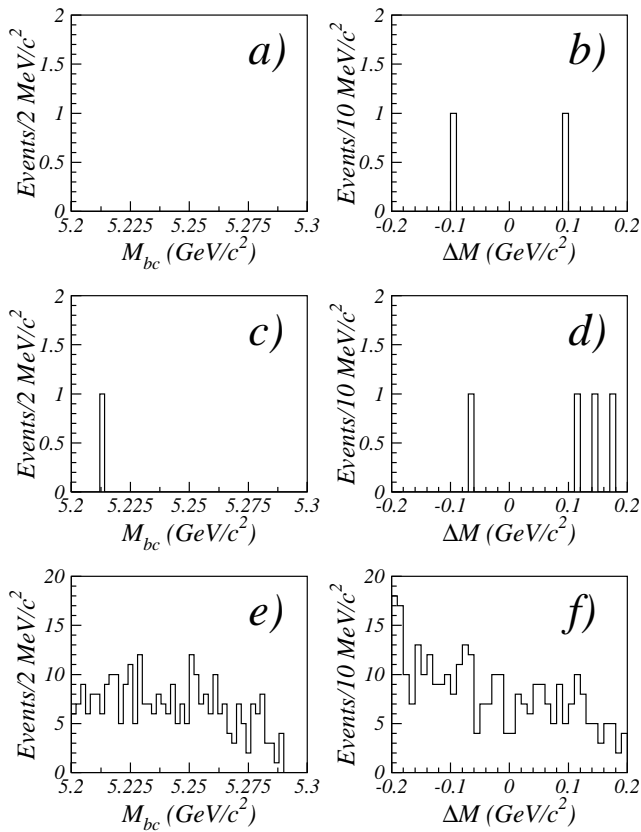}
\caption{ 
Cross-checks of the signal. (a) and (b) are the $M_{\rm bc}$ and $\Delta M$ distributions for the on-resonance data with the inverted requirement on the normalized Fox-Wolfram moment $R_2$ (see text for details).
(c) and (d) are the $M_{\rm bc}$ and $\Delta M$ distributions for the continuum data.
(e) and (f) are the $M_{\rm bc}$ and $\Delta M$ distributions for the primary vertex protons with inverted identification. No peaking structure in the signal region is present in any of the cross-check analyses.}
\label{3}
\end{figure}

The $M_{\rm bc}$ and $\Delta M$ distributions for the $\overline{B}{}^0 \ra D_s^+ \Lambda \overline{p}$ candidates are shown in Fig.\,\ref{2} (a) and (b), respectively,  where all three $D_s$ modes are combined. 
We require $M_{\rm bc} > 5.272$\,GeV/$c^2$ ($|\Delta M|<0.025$\,GeV/$c^2$) for the $\Delta M$ ($M_{\rm bc}$) projection. We found that after applying all the selection requirements, there are no events counted repeatedly in the $M_{\rm bc}$ and $\Delta M$ distributions. 
The hatched histograms in Fig.\,\ref{2} (a) and (b) show normalized $D_s$ mass sidebands where no peaking structures are evident. The superimposed curves are the results of a simultaneous two-dimensional binned maximum likelihood fit (with common branching fraction as a constraint) to the three $\Delta M$ versus $M_{\rm bc}$ distributions (for the three $D_s$ channels). 

To describe the signal we use Gaussians with means and widths fixed to the values obtained from MC. The backgrounds in $M_{\rm bc}$ and $\Delta M$ are parameterized by a first-order polynomial and an ARGUS function \cite{ARGUS_function}, respectively. 
The fit gives a statistical significance of $6.6 \sigma$ for the signal, where the statistical significance is defined as $\sqrt{-2 \ln(L_0/L_{\text{max}})}$, where $L_0$ and $L_{\text{max}}$ are the likelihoods with the signal fixed at zero and at fitted value, respectively. 
The region $\Delta M < -0.08$\,GeV/$c^2$ is excluded from the fit to avoid possible contributions from the $\overline{B}{}^0 \ra D_s^{*+} \Lambda \overline{p}$, $D_s^{*+} \ra D_s^+ \gamma$ and $B^- \ra D_s^{+}\Lambda\overline{p}\pi^-$ decays, where the soft $\gamma$($\pi^-$) is undetected. The choice of the fitting range is taken into account in the systematic error.
The results of the fit applied 
for the three $D_s^+$ modes separately are shown in Table\,\ref{2D}.

We select events in the $B$-signal region of $|\Delta M|<0.025$\,GeV/$c^2$ and
 $M_{\rm bc}>5.272$\,GeV/$c^2$ for the three $D_s^+$ modes and examine the two-baryon
 invariant mass distribution (Fig.\,\ref{2} (c)). 
We see apparent threshold peaking behavior 
of this distribution, while the $B$-sideband \cite{sideband} distribution is smooth. 
Such a threshold peaking behavior seems intrinsic to all multi-body baryonic $B$ decays. The invariant mass distribution for $D_s^+$ and $\Lambda$ and for corresponding $B$-sideband is represented as well (Fig.\,\ref{2} (d)). 
Some peaking behavior is also seen in this distribution, which could 
arise from some new excited charm baryon. Firm conclusions, however, 
cannot yet be drawn on either peaks because of limited statistics.

As a cross-check 
we analyze the on-resonance data with the inverted requirement 
on the normalized Fox-Wolfram moment $R_2>0.5$ (Fig.\,\ref{3} (a), (b)) and 
off-resonance data sample (Fig.\,\ref{3} (c), (d)) and find no candidates in 
the $B$-signal region. The distributions for the primary vertex protons 
with an inverted particle identification requirement \cite{inverted} do not peak in the signal region (Fig.\,\ref{3}  (e), (f)), demonstrating that the selected $\overline{B}{}^0 \ra D_s^+ \Lambda \overline{p}$ candidates contain real protons.

Table\,\ref{2D} 
summarizes the results of the fits, the reconstruction efficiencies including the ${\cal B}(D_s^+ \ra \phi\pi^+)$, ${\cal B}(\phi \ra K^+K^-)$, ${\cal B}(D_s^+ \ra \overline{K}{}^{*0}K^+)$, ${\cal B}(\overline{K}{}^{*0} \ra K^-\pi^+)$, ${\cal B}(D_s^+ \ra K_S^0 K^+)$, ${\cal B}(K^0_S\ra\pi^+\pi^-)$, ${\cal B}(\Lambda \ra p \pi^-)$ branching fractions, statistical significance of the signals and extracted branching fractions. Here we assume equal fractions of charged and neutral $B$ mesons produced in $\Upsilon(4S)$ decays.

The major sources of systematic error are the uncertainties in the tracking efficiency 6\% (1\% per track), 12\% in the charged particle identification efficiency (1\% for pion, 2\% for kaon, 3\% for proton), 5\% for $\Lambda$ finding, 3\% for efficiency estimation due to MC statistics and 5\% for the error due to choice of the fitting procedure. These contributions are combined in quadrature resulting in a total systematic error of 16\%. We also take into account a third error due to the uncertainty in ${\cal B}(D_s^+\rightarrow\phi\pi^+)$ that is 14\%.

In summary, we report the first observation of the  decay $\overline{B}{}^0 \ra D_s^+ \Lambda \overline{p}$ with a branching fraction of $(2.9 \pm 0.7 \pm 0.5 \pm 0.4)\times 10^{-5}$, where the
first error is statistical, the second is systematic, and the third arises from 
uncertainty in the branching  fraction of $D_s^+\rightarrow\phi\pi^+$. The
statistical significance is $6.6\,\sigma$. 
This charmful decay can occur via the creation of an $s\bar{s}$ pair or from the more copious 
$\overline{B} \to D N \overline{N}$ modes with $(DN)^+ \to D_s^+ \Lambda$ 
rescattering in the final state \cite{BELLE5,CLEO2,BABAR}. In the future, this decay 
mode can be used for $CP$ asymmetry studies.

We thank the KEKB group for excellent operation of the
accelerator, the KEK cryogenics group for efficient solenoid
operations, and the KEK computer group and
the NII for valuable computing and Super-SINET network
support.  We acknowledge support from MEXT and JSPS (Japan);
ARC and DEST (Australia); NSFC and KIP of CAS (China); 
DST (India); MOEHRD, KOSEF and KRF (Korea); 
KBN (Poland); MES and RFAAE (Russia); ARRS (Slovenia); SNSF (Switzerland); 
NSC and MOE (Taiwan); and DOE (USA).


\begin{thebibliography}{99}

\bibitem{BELLE1} Belle Collaboration, N. Gabyshev {\it et al.,} \prl {\bf 90}, 121802 (2003).
\bibitem{BELLE2} Belle Collaboration, M.-Z. Wang {\it et al.,} \prl {\bf 90}, 201802 (2003).
\bibitem{BELLE3} Belle Collaboration, K. Abe {\it et al.,} \prl {\bf 88}, 181803 (2002).
\bibitem{BELLE4} Belle Collaboration, N. Gabyshev {\it et al.,} \pr D {\bf 66}, 091102 (2002).
\bibitem{BELLE5} Belle Collaboration, K.~Abe {\it et al,} \prl {\bf 89}, 151802 (2002).
\bibitem{CLEO1} CLEO Collaboration, S. A. Dytman {\it et al.,} \pr D {\bf 66}, 091101 (2002).
\bibitem{CLEO2} CLEO Collaboration, S. Anderson {\it et al.,} \prl {\bf 86}, 2732 (2001).
\bibitem{BELLE6} Belle Collaboration, M.-Z. Wang {\it et al.,} \prl {\bf 92}, 131801 (2004).
\bibitem{BELLE7} Belle Collaboration, K. S. Park, H. Kichimi {\it et al.,} \pr D {\bf 75}, 011101 (2007).
\bibitem{BELLE8} Belle Collaboration, M.-Z. Wang, Y.-J. Lee {\it et al.,} Belle preprint 2007-19, arXiv: 0704.2672[hep-ex].
\bibitem{HOU1} C.-K. Chua, W.-S. Hou, S.-Y. Tsai, \pr D {\bf 65}, 034003 (2002).
\bibitem{HOU3} C.-K. Chua, W.-S. Hou, S.-Y. Tsai, \pl B {\bf 528}, 233 (2002).
\bibitem{ROSNER} J. L. Rosner, \pr D {\bf 68}, 014004 (2003).
\bibitem{KERBIKOV} B. Kerbikov, A. Stavinsky and V. Fedotov, \pr C {\bf 69}, 055205 (2004).
\bibitem{HAIDEN} J. Haidenbauer, Ulf-G. Meissner and A. Sibirtsev, \pr D {\bf 74}, 017501 (2006).
\bibitem{ENTEM} D. R. Entem and F. Fernandez, \pr D {\bf 75}, 014004 (2007). 
\bibitem{THRESHOLD} H. Kichimi, Nucl. Phys. B Proc. Suppl. {\bf 142}, 197 (2005).
\bibitem{HOU2} W.-S. Hou, A. Soni, \prl {\bf 86}, 4247 (2001).
\bibitem{DALITZ} BABAR Collaboration, B. Aubert {\it et al.,} \pr D {\bf 72}, 051101 (2005).
\bibitem{BELLE9} Belle Collaboration, M.-Z. Wang {\it et al.,} \pl B {\bf 617}, 141 (2005).

\bibitem{kekb} S.~Kurokawa and E.~Kikutani, Nucl. Instr. Methods Phys. Res., Sect. A {\bf 499}, 1 (2003), and other papers included in this volume.

\bibitem{belle_detector} Belle Collaboration, A.~Abashian {\it et al.,} Nucl. Instr. Methods Phys. Res., Sect. A {\bf 479}, 117 (2002). 

\bibitem{svd2} Z.~Natkaniec {\it et al.} (Belle SVD2 Group), Nucl. Instr. and Meth. A {\bf 560}, 1 (2006).

\bibitem{montecarlo} R.~Brun {\it et al.}, GEANT 3.21, CERN DD/EE/84-1, 1984. 

\bibitem{ID}
Charged kaons are required to satisfy 
${\cal L}(K)/({\cal L}(K)+{\cal L}(\pi))>0.6$.
Charged pions are required to satisfy 
${\cal L}(\pi)/({\cal L}(K)+{\cal L}(\pi))>0.1$.
Protons are required to satisfy 
${\cal L}(p)/({\cal L}(K)+{\cal L}(p))>0.6$ and 
${\cal L}(p)/({\cal L}(\pi)+{\cal L}(p))>0.6$. 
Here ${\cal L}(K/\pi/p)$ is the 
particle identification likelihood for the $K/\pi/p$ hypotheses.
The above requirements have efficiencies of more than $95\%$ for pions, kaons and protons, respectively, from $\overline{B}{}^0 \ra D_s^+ \Lambda \overline{p}$ decays. The probability for each particle species to be misidentified as one of the other two is less than $5\%$.  


\bibitem{PDG} W.-M.~Yao {\it et al.,} (Particle Data Group) J.~ Phys. G {\bf 33}, 1 (2006).

\bibitem{fox} G.~C.~Fox and S.~Wolfram \prl {\bf 41}, 1581 (1978).
\bibitem{DM1} Belle Collaboration, S.~L.~Zang {\it et al.,} \pr D {\bf 69,} 017101 (2004).
\bibitem{DM2} Belle Collaboration, N.~Gabyshev {\it et al.,} \prl {\bf 97}, 202003 (2006). 
\bibitem{ARGUS_function} ARGUS Collaboration, H. Albrecht {\it et al.,} \pl B {\bf 241}, 278 (1990).
\bibitem{sideband} 
The $B$-sideband is defined as a region of $M_{\rm bc}>5.2$\,GeV/$c^2$ and $-0.08$\,GeV/$c^2<\Delta M<0.12$\,GeV/$c^2$ excluding the $B$-signal region.
\bibitem{inverted}
Protons are required to satisfy inverted likelihood ratio requirement 
${\cal L}(p)/({\cal L}(K)+{\cal L}(p))<0.6$. 
\bibitem{BABAR} BABAR Collaboration, B.~Aubert {\it et al.,} \pr D {\bf 74}, 051101 (2006).



\end{thebibliography}
\end{document}